\documentclass[floatfix,prl,showpacs,twocolumn,aps]{revtex4}
\usepackage{graphicx}

\newcommand{\ket}[1]{\left|#1\right\rangle}
\newcommand{\bra}[1]{\left\langle#1\right|}
\begin{document}

\title{Cavity QED in superconducting circuits:
susceptibility at elevated temperatures}

\author{Ileana Rau}
\altaffiliation[Present address:]{Physics Department, Stanford University}
\author{G\"oran Johansson}
\author{Alexander Shnirman}
\affiliation{Institut f\"ur Theoretische Festk\"orperphysik,
Universit\"at Karlsruhe, D-76128 Karlsruhe, Germany}

\begin{abstract}
We study the properties of superconducting electrical circuits,
realizing cavity QED. In particular we explore the limit
of strong coupling, low dissipation, and elevated temperatures relevant
for current and future experiments. We concentrate on the cavity
susceptibility as it can be directly experimentally addressed, i.e., as
the impedance or the reflection coefficient of the cavity.
To this end we investigate the dissipative Jaynes-Cummings model
in the strong coupling regime at high temperatures.
The dynamics is investigated within the Bloch-Redfield formalism.
At low temperatures, when only the few lowest levels are occupied the
susceptibility can be presented as a sum of contributions
from independent level-to-level transitions. This corresponds
to the secular (random phase) approximation in the Bloch-Redfield
formalism.
At temperatures comparable to and higher than
the oscillator frequency, many transitions become important and
a multiple-peak structure appears. We show that in this regime
the secular approximation breaks down, as soon as the peaks start
to overlap. In other words, the susceptibility is no longer a sum
of contributions from independent transitions.
We treat the dynamics of the system numerically by exact
diagonalization of the Hamiltonian of the qubit plus up to 200 states
of the oscillator. We compare the results obtained with and without
the secular approximation and find a qualitative discrepancy
already at moderate temperatures.
\end{abstract}
\pacs{85.25.-j, 42.50.Pq, 85.25.Cp, 03.67.Lx}
\maketitle

\section{Introduction}

There is, currently, a substantial activity in the research into the physics of
Josephson qubits. 
In particular many 
groups~\cite{Buisson_QED,Blais_Schoelkopf_et_al,Jena_Rabi,Irish_Schwab}
became interested in studying the systems composed of Josephson 
qubits and harmonic
oscillators, e.g., microwave cavities, mechanical resonators etc.
Thus the field starts to resemble quantum optics where atoms in cavities have
been investigated for many years.
On one hand there are many results which can be simply ``translated'' from
the ``language'' of quantum optics to the ``language'' of solid state physics.
On the other hand there are specific properties of the solid state devices
that might require further research.

In this paper we study the dynamics of a two-level system (qubit) and a cavity
at resonance. In quantum optics this regime is the most studied and interesting
one. The corresponding Jaynes-Cummings model (for a review see 
Ref.~\onlinecite{Shore_Knight_Review})
has been widely investigated in the literature. Experimentally, however, 
the strong coupling
regime is difficult to achieve and it is also a challenge to keep 
the atom in the cavity
for a long time. In optical cavities the strong coupling regime was achieved only
a decade ago~\cite{Kimble_strong_QED}. Rydberg atoms in superconducting
cavities~\cite{Raimond_Haroche_RMP} provide the strong coupling regime and one
can even perform quantum gates during the time of flight of the atom through the
cavity.

In solid state devices, e.g., a Josephson qubit resonantly coupled
to a damped strip-line superconducting
cavity~\cite{Blais_Schoelkopf_et_al}, the "atom" is permanently
placed in the cavity. This should simplify the time constraints.
Also, the strong coupling limit seems to be possible. Thus these
circuits constitute very promising setups for exploring the
strong coupling limit of cavity QED. Compared to optical cavities,
one of the differences is the finite temperature of
solid state devices. Moreover, even if the nominal
temperature of the refrigerator is much lower than the cavity's
energy splitting, the hot photons can arrive via the leads connecting
to the controlling circuits. Thus the elevated temperature regime is
of substantial interest. We are aware of only one article
addressing the finite temperature case,
namely~Ref.~\onlinecite{Cirac_Dissipative_JCM}. In this paper we expand
the domain of parameters as compared with
Ref.~\onlinecite{Cirac_Dissipative_JCM}. Namely we consider the case
when both the qubit's and the cavity's dissipative rates are much
smaller than the cavity-qubit coupling (Rabi-frequency) and both
the cavity and the qubit are coupled to the finite temperature
baths. Moreover we focus on the correlation functions of the
cavity, e.g., cavity susceptibility. In solid state systems this
quantity is particularly convenient to measure as it is related to
the impedance of the strip-line cavity. The temperature-dependence
of this impedance constitutes a direct tool for probing the number
of photons in the cavity.

At low temperatures the dynamics is very simple. Only a
few states are occupied and the susceptibility may be presented as a
sum of Lorentzians, corresponding to a few transitions allowed
from these states. E.g., at zero temperature only two transitions
are relevant and the resonant peak of the uncoupled cavity is Rabi
split by the qubit to two peaks. At high temperatures the
frequencies of different allowed transitions are densely packed
near the oscillator's frequency. This situation is called
Liouvillian degeneracy as, formally, different modes of the
Liouville evolution operator are almost degenerate~\cite{Alec_Liouville}. We show that
the secular approximation, widely used within the Bloch-Redfield
formalism, fails due to the Liouvillian degeneracy. The
insufficiency of the secular approximation was already noticed in,
e.g., Ref.~\onlinecite{Haenggi_Rectification_RWA}. In this situation one
has to take more elements of the Redfield tensor into account than
required by the secular approximation. On the other hand the
optical master equation takes all the necessary elements into
account and thus produces correct results.

\section{Experimental motivation}
We follow the recent proposal, presented in Ref.~\onlinecite{Blais_Schoelkopf_et_al},
in which a superconducting charge qubit, coupled capacitatively to a
cavity formed in a coplanar waveguide,
is shown to be a favorable system for reaching the strong coupling regime
of cavity QED.
Typical parameters are cavity resonance/atom transition
frequency of $\omega_0/2\pi=10$ GHz, a vacuum Rabi-frequency of 
$g/2\pi=100$ MHz,
a cavity lifetime of $1/\kappa=160$ ns (quality factor $Q=\omega_0/\kappa=10^4$), 
and an atom lifetime $1/\gamma=2 \mu$s.
The system is measured by detecting absorption and phase shift of
microwaves sent through the waveguide.
In this paper we calculate the linear response absorption
of the system. Motivated by the
experimental parameters we focus
on the regime of strong coupling $(\kappa,\gamma \ll g)$, and also
assume that cavity dissipation dominates over atom dissipation
$(\gamma < \kappa)$. We note that the photon temperature in the cavity
may be much higher than the base temperature of the cryostat,
due to coupling to room temperature sources through the waveguide.
By measuring the absorption spectra one may determine the photon
temperature of the cavity.

\section{The Jaynes-Cummings model}
\begin{figure}
\centerline{\hbox{
\includegraphics[width=0.9\columnwidth]{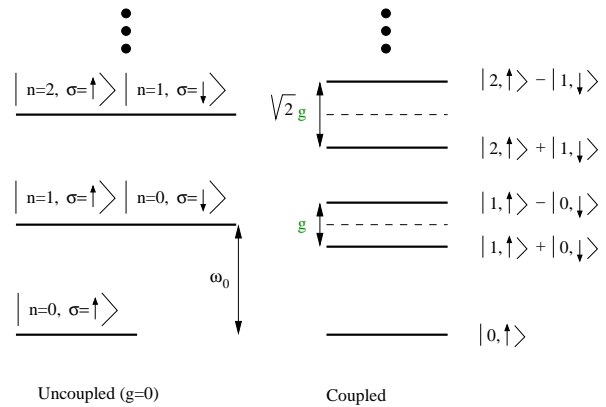}
}}
\caption[]  {\label{Fig:Spectrum}
Spectrum of the Jaynes-Cummings model at resonance.}
\end{figure}

Assuming the qubit to be at the degeneracy point 
\cite{Saclay_Manipulation_Science,Blais_Schoelkopf_et_al}, 
and for the moment neglecting dissipation, we arrive at the Jaynes-Cummings model,
described by the following Hamiltonian
\begin{equation}
\label{JCham}
H_{JC}=-\frac{1}{2}\omega_{qb} \sigma_z
+ \omega_{osc}\,a^\dag a
+ \frac{g}{2}\,\sigma_x \left(a+a^\dag\right)
\ ,
\end{equation}
where $\sigma$ operates on atom/qubit while
$a, a^\dag$ are ladder operators of the cavity/oscillator.
We consider the resonant regime when $\omega_{osc}=\omega_{qb}=\omega_0$.
The system's spectrum for $g\ll \omega_0$ is shown in Fig.~\ref{Fig:Spectrum}.
It is obtained by, first, analyzing the spectrum for $g=0$ (left
side) and, then, lifting the degeneracies by the coupling $g$
(right side). Assuming $g \ll \omega_0$ we can take into account
the coupling term only when it lifts the degeneracies in the
spectrum. The ground state of the uncoupled system is
$\ket{g}=\ket{\uparrow,0}$, which is non-degenerate. All
other states are doubly degenerate, i.e. the states
$\ket{\uparrow,n}$ and $\ket{\downarrow,n-1}$ form degenerate
doublets for each $n \ge 1$. Every such doublet is split by the
interaction. We can thus define the "bonding" and the
"anti-bonding" states $|b/a,n\rangle \equiv 2^{-1/2}
\left(\ket{\uparrow,n} \pm \ket{\downarrow,n-1}\right)$. The
energies of these states are $E_{b/a,n} = n\omega_0 \mp
\sqrt{n}\,g/2$.

\section{Oscillator/Cavity impedance}

In ordinary cavity QED the atom susceptibility is usually probed.
For the qubit-oscillator system the susceptibility of the
oscillator (cavity) is the most easily measured quantity.
It is defined as
\begin{equation}
\chi(t) = i\theta(t)\langle [q(t),q(0)]_{-}\rangle \ ,
\end{equation}
where $q\equiv a^{\dag}+a$. The Fourier transform is
$\chi(\omega)=\chi'(\omega)+i\chi''(\omega)=\int dt \chi(t)
e^{i\omega t}$. The imaginary part of the susceptibility is
proportional to the dissipative real part of the impedance, i.e.,
$\chi''(\omega) \propto {\rm Re}\,Z(\omega)$. For a closed system
with quantum levels $\ket{i}$, with stationary occupation probabilities
$\rho_i$, it can be presented as
\begin{eqnarray}
\chi''(\omega) &=& \pi
\sum_{i,f}\,\rho_i\,|\bra{i}q\ket{f}|^2\,\nonumber\\
&\times&\big(\delta(\omega-\omega_{fi})-
\delta(\omega+\omega_{fi})\big)\ ,
\end{eqnarray}
where $\omega_{fi}\equiv (E_f-E_i)/\hbar$. Thus it is a series of
delta-peaks, corresponding to the allowed transitions.

\begin{figure}
\centerline{\hbox{
\includegraphics[width=0.9\columnwidth]{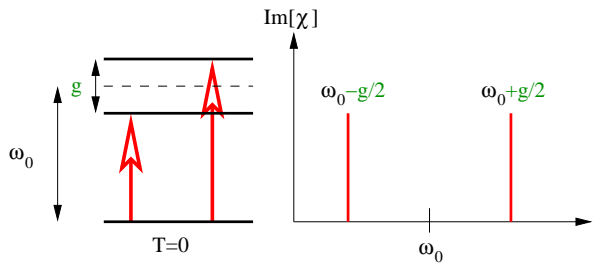}
}}
\caption[]  {\label{Fig:RabiSplit}
Transitions (left) resulting in a Rabi
splitting of the cavity susceptibility (right), at $T=0$.}
\end{figure}

At $T=0$ only the ground state is occupied and the two allowed
transitions have frequencies $\omega_{g;\,b,1} = \omega_0 - g/2$
and $\omega_{g;\,a,1} = \omega_0 + g/2$ and the matrix elements
$|\bra{g}q\ket{b,1}|=|\bra{g}q\ket{a,1}|=1/\sqrt{2}$. Thus the
uncoupled oscillator's peak at $\omega=\omega_0$ is split as shown
in Fig.~\ref{Fig:RabiSplit}. This is called Rabi-splitting.

\begin{figure}
\centerline{\hbox{
\includegraphics[width=0.9\columnwidth]{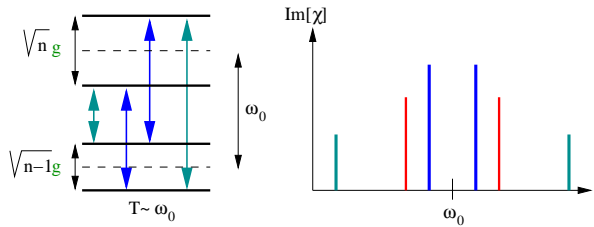}
}}  \caption[]  {\label{Fig:FiniteTImp}
Transitions (left) and cavity susceptibility (right) at $T>0$.
The central (blue) arrows correspond
to "bonding-bonding" and "anti-bonding-anti-bonding" transitions,
giving rise to the central two peaks in the spectrum.
The outer (blue-green) arrows correspond to "bonding-anti-bonding"
transitions, giving the outer two peaks in the spectrum. }
\end{figure}

At $T>0$ higher energy states are occupied. The transitions are
allowed only between neighboring doublets. There are two classes
of allowed transitions. The "bonding-bonding" or
"anti-bonding-anti-bonding" transitions correspond to the
transition frequencies
\begin{eqnarray}
\label{bb_freq}
\omega_{b,n-1;\,b,n} = \omega_0 -
{\frac{g}{2}}{\left(\sqrt{n}-\sqrt{n-1}\right)} \ ,\\
\omega_{a,n-1;\,a,n} = \omega_0 +
{\frac{g}{2}}{\left(\sqrt{n}-\sqrt{n-1}\right)} \ ,
\end{eqnarray}
with the matrix elements
\begin{eqnarray}
&&|\bra{b,n-1}q\ket{b,n}|=|\bra{a,n-1}q\ket{a,n}|=\nonumber \\&&=
\frac{1}{2}\left(\sqrt{n}+\sqrt{n-1}\right) \ .
\end{eqnarray}
They form the first class of transitions which all are positioned
inside of the zero temperature Rabi-splitting. Analogously, the
"bonding-anti-bonding" transitions with frequencies
\begin{eqnarray}
\omega_{b,n-1;\,a,n} = \omega_0 +
{\frac{g}{2}}{\left(\sqrt{n}+\sqrt{n-1}\right)} \ ,\\
\omega_{a,n-1;\,b,n} = \omega_0 -
{\frac{g}{2}}{\left(\sqrt{n}+\sqrt{n-1}\right)} \ ,
\end{eqnarray}
and matrix elements
\begin{eqnarray}
\label{ba_mat_elements}
&&|\bra{b,n-1}q\ket{a,n}|=|\bra{a,n-1}q\ket{b,n}|=\nonumber \\&&=
\frac{1}{2}\left(\sqrt{n}-\sqrt{n-1}\right)
\end{eqnarray}
are all positioned outside the Rabi-splitting. Both classes of
transitions are shown in Fig.~\ref{Fig:FiniteTImp}.

\section{Cavity dissipation}

Due to the dissipation the delta functions should be widened to
Lorentzians. At zero temperature the widths of the peaks are estimated as
$\delta\omega = \omega_0/(2Q)$ (the factor $2$ is due to the
reduced matrix element for each transition as compared to the
original oscillator's transition).

The problem is getting the line-shape at finite (relatively
high) temperatures and high values of $Q$. Indeed, when
temperature becomes of order $\omega_0$ more transitions become
available. To estimate the heights of the peaks we note that the
occupation probabilities of the doublets are estimated as $\rho_n
\approx Z \exp(-n\omega_0/T)$. The matrix elements for the
"bonding-bonding" or "anti-bonding-anti-bonding" transitions grow
as $\propto \sqrt{n}$, while those for "bonding-anti-bonding"
transitions decay as $\propto 1/\sqrt{n}$. Thus we obtain a
"Poisson" distribution for the "bonding-bonding" or
"anti-bonding-anti-bonding" peaks' heights with the maximum at $n
\approx T/\omega_{0}$. The heights of the
"bonding-anti-bonding" peaks decay with $n$. Therefore the highest
peaks are situated at $\omega \approx \omega_0 \pm
\sqrt{\omega_0/T}\,g/4$.

The spacing between the dominant peaks ($\sim \frac{g}{2}\,\sqrt{\frac{\omega_0}{T}}$)
thus decrease with temperature, while their
width ($\sim \frac{2T^2}{Q\omega_0}$) 
increase with temperature.
Around the cross-over temperature $T_c=\omega_0(gQ/4\omega_0)^{2/5}$
the peaks start to overlap.

\begin{figure}
\centerline{\hbox{
\includegraphics[width=0.3\columnwidth]{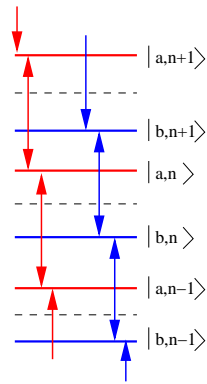}
}}
\caption[]  {\label{Fig:HighTempSpec}
At high temperatures $T\gg\omega_0$ the ``bonding'' (blue) and
``anti-bonding'' (red) states form two almost equidistant spectrums.
The non-harmonicity decreases as $\sqrt{n+1}-\sqrt{n} \approx 1/2\sqrt{n}$
for large $n$.}
\end{figure}

Thus, as the temperature grows, we have a transition from a quite 
"un-harmonic" spectrum without Liouvillian symmetry, where all transition
frequencies are different, to a spectrum which resembles that of two
uncoupled linear oscillators, see Fig. \ref{Fig:HighTempSpec}. 
The widths of the peaks in these two cases
behave qualitatively different. In the first (un-harmonic) case the widths
grow with temperature, while in the second (harmonic oscillator)
case they do not.

\section{Qubit/atom dissipation}
The qubit/atom is also subjected to dissipation, which will add
to the peak widths in the susceptibility. At the degeneracy point
of the qubit the longitudinal noise (coupled to $\sigma_z$) is suppressed
to first order~\cite{Saclay_Manipulation_Science}. 
The transverse noise (coupled to $\sigma_x$) will
induce transitions in the qubit, characterized by the rate $1/\gamma$.  
In terms of the Jaynes-Cummings eigenstates the qubit dissipation
cause similar transitions as the cavity dissipation. A major difference
is that the matrix elements for these transitions are independent
of the oscillator state $n$. Thus, as long as the two sources of dissipation
are of the same order of magnitude,
the high temperature susceptibility will be determined by the
cavity dissipation.

\section{Bloch-Redfield formalism}

We model the cavity/atom dissipation by coupling an observable $X/Y$ of a thermal bath 
to the $q/\sigma_x$-coordinate of the cavity/atom:
\begin{equation}
H=H_{JC}+ q X + \sigma_x Y + H^X_{\rm bath} + H^Y_{\rm bath}
\ .
\end{equation}
One popular choice of a bath is the harmonic oscillator 
one~\cite{Vernon63,Caldeira_Leggett_PRL81}. In this case
$H_{\rm bath}=\sum_i \omega_i b^\dag_i b_i$ and 
$X=\sum_i \lambda_i (b_i+b^\dag_i)$. However, as we 
are assuming weak dissipation and doing the lowest order 
calculation, the precise nature of the bath is unimportant
and only the bath correlator $\langle X(t)X(0)\rangle$ matters.

The Bloch-Redfield 
equation~\cite{Bloch_Derivation,Redfield_Derivation} 
is a kinetic equation for the reduced (qubit+cavity) density
matrix:
\begin{equation}
\dot\rho_{mn}(t)+ i \omega_{mn} \rho_{mn}(t) =\sum_{m',n'}
R_{mnm'n'}\ \rho_{m'n'}(t) \ ,
\label{BReq}
\end{equation}
where the Bloch-Redfield tensor is given by
\begin{eqnarray}
R_{mnm'n'}&=&\lambda_{m'mnn'}+\lambda^*_{n'nmm'}
\nonumber \\
&-&
\sum_k \delta_{mm'}\lambda_{nkkn'}+\delta_{nn'}\lambda^*_{mkkm'}
\ ,
\end{eqnarray}
and
\begin{eqnarray}
\label{lambda_def}
\lambda_{m'mnn'}&=&\bra{m'}q\ket{m}\bra{n}q\ket{n'} L_X(\omega_{nn'}) \nonumber\\
&+& \bra{m'}\sigma_x\ket{m}\bra{n}\sigma_x\ket{n'} L_Y(\omega_{nn'})
\ ,
\end{eqnarray}
where $L_{X}(\omega)$ is the Laplace transform of the cavity bath correlator
\begin{equation}
L_X(\omega)=\int_0^\infty dt e^{-i\omega t}\langle X(t)X(0) \rangle
\ .
\end{equation}
Introducing the Fourier image of the unsymmetrized correlator 
$\langle X^2_{\nu} \rangle 
\equiv \int dt \langle X(t)X \rangle\,e^{i\nu t}$, we obtain
\begin{equation}
L_X(\omega) = -i \int \frac{d\nu}{2\pi}\, \frac{\langle X^2_{\nu} \rangle}{\nu+\omega-i0}
\ .
\end{equation}
Thus for the real part of $L_X$ which determines the transition rates 
we obtain ${\rm Re}\,L_X(\omega)=(1/2)\,\langle X^2_{\nu=-\omega} \rangle$.
The imaginary part of $L_X$ is responsible for the energy shifts (Lamb shift).
$L_Y(\omega)$ of the qubit bath is defined and treated analogously.

Considering a high quality cavity ($Q\gg1$), and low qubit dissipation ($\gamma \ll \omega_0$),
we limit ourselves to single photon exchange with the baths.
The possible transition energies in Eq.~(\ref{lambda_def}) are
$\omega_{nn'}=\pm\omega_0+O(g)$, (see Eqs.~(\ref{bb_freq})-(\ref{ba_mat_elements})).
Furthermore we assume that the baths have no structure
on the scale of the qubit-oscillator coupling $g$,
leaving only one relevant parameter for relaxation
$\langle X^2_{\omega_0} \rangle =\kappa/(1-e^{-\omega_0/T_\kappa})$,
$\langle Y^2_{\omega_0} \rangle =\gamma/(1-e^{-\omega_0/T_\gamma})$,
and one for excitation 
$\langle X^2_{-\omega_0} \rangle=e^{-\omega_0/T_\kappa}\langle X^2_{\omega_0} \rangle$, 
$\langle Y^2_{-\omega_0} \rangle=e^{-\omega_0/T_\gamma}\langle Y^2_{\omega_0} \rangle$,
for each bath.

\subsection{The secular approximation}
Looking for the slow dynamics of our system, on the time-scale
given by the dissipation  ($1/\kappa, 1/\gamma$), 
we rewrite eq.~(\ref{BReq}) in the rotating frame
\begin{equation}
\dot{\tilde{\rho}}_{mn}(t)=\sum_{m',n'}
R_{mnm'n'}\ \tilde{\rho}_{m'n'}(t) e^{i(\omega_{mn}-\omega_{m'n'})t} \ ,
\end{equation}
where $\tilde{\rho}_{mn}(t)=\rho_{mn}(t)e^{i\omega_{mn}t}$ evolves
slowly in time.

When there are no Liouvillian degeneracies the phase
$(\omega_{mn}-\omega_{m'n'})t$ in the above expression
rotates with a frequency of the order of $\omega_0 \gg \kappa, \gamma$,
except when $m=m'$ and $n=n'$, or $m=m'$ and $n=n'$.
The {\it secular} approximation is a random-phase type of approximation,
keeping {\it only} the corresponding elements of Redfield tensor,
$R_{mnmn}$ and $R_{mmnn}$.

%

Within the secular approximation the Bloch-Redfield
equation separates into a master equation
governing the occupation numbers:
$\dot{\rho_{nn}}(t)=R_{nnmm} \rho_{mm}(t)$, and a simple
exponential decay of all off-diagonal elements of the density matrix:
$\rho_{mn}(t)=\rho_{mn}(0)e^{(-i\omega_{mn}+R_{mnmn})t}$.
In the susceptibility we find that the weights of the peaks
are given by the steady-state occupation numbers, determined
by the transition rates ($R_{nnmm}$), while the peak widths are given
by the dephasing rates ($R_{mnmn}$).

\subsection{Liouvillian degeneracy}
When difference between two transition
frequencies $\omega_{mn}-\omega_{m',n'}$
($m \neq m'$ and $n\neq n'$)
becomes smaller than the transitions' widths
$\propto (\kappa, \gamma)$ the peaks start to overlap.
In this case there is no justification for the secular
approximation, and more elements of the Redfield tensor must
be retained.

In the lowest order (single-photon transitions) 
the allowed transition (photon) frequencies are
$\pm\omega_0+O(g)$, giving two possible values for
$|\omega_{mn}-\omega_{m'n'}|=\{0,2\omega_0\}+O(g)$.
Looking for slow dynamics we neglect the elements $R_{mnm'n'}$
corresponding to the Liouvillian modes with 
$|\omega_{mn}-\omega_{m'n'}|=2\omega_0+O(g)$.
This is equivalent to a rotating-wave-approximation 
in the coupling to the bath. We keep, however, the elements $R_{mnm'n'}$
corresponding to the Liouvillian modes with 
$|\omega_{mn}-\omega_{m'n'}|=O(g)$. That is, 
grouping the elements of the density matrix $\rho_{mn}$ according
to the energy difference $E_m-E_n=M \omega_0+O(g)$, we 
now couple only the elements with similar $M$,
but leave the elements with different $M$ uncoupled.
We also note that the Bloch-Redfield equation with this choise
of the elements $R_{mnm'n'}$ is
equivalent to the "Quantum Optics" master equation:
\begin{eqnarray}
\dot\rho&=&-i[H_{JC},\rho]+ 
\kappa N_\kappa (2a^\dag\rho a-a a^\dag\rho-\rho a a^\dag)\nonumber \\
&+&\kappa (N_\kappa+1)(2a\rho a^\dag-a^\dag a\rho-\rho a^\dag a)\nonumber \\
&+&\gamma N_\gamma (2\sigma^+\rho \sigma^- 
-\sigma^-\sigma^+ \rho-\rho \sigma^-\sigma^+)\nonumber \\
&+&\kappa (N_\gamma+1)(2\sigma^-\rho\sigma^+ -\sigma^+\sigma^-\rho-\rho\sigma^+\sigma^- )
\end{eqnarray}
where $N_{\kappa/\gamma}=1/(e^{\omega_0/T_{\kappa/\gamma}}-1)$ are the average occupation 
numbers of each bath at frequency $\omega_0$.

\section{Numerical approach}
In our numerical studies we take into account the
lowest $N$ states of the harmonic oscillator, and
then check for convergence by increasing $N$.
We work in the eigenbasis of the Jaynes-Cummings Hamiltonian
(Eq.~\ref{JCham}), which we obtain by numerical diagonalization.

We then rewrite the Bloch-Redfield equation (Eq.~(\ref{BReq}))
in a matrix form, i.e.
\begin{equation}
\dot\rho_{[mn]}(t)+ i \omega_{[mn][mn]} \rho_{[mn]}(t) =
\sum_{[m'n']} R_{[mn][m'n']}\ \rho_{[m'n']}(t) \ ,
\label{BR_mat_eq}
\end{equation}
where $\hat{\rho}(t)=\rho_{[mn]}(t)$ now is a column vector of length $(2N)^2$
and $\hat{R}=R_{[mn][m'n']}$ is a matrix of size $(2N)^2 \times (2N)^2$,
and $\hat{\omega}=\omega_{[mn][mn]}$ is a diagonal matrix of the same size.
The solution can then be written
\begin{equation}
\label{BRsol}
\hat{\rho(t)}=e^{(-i\hat{\omega}+\hat{R})t} \cdot \hat{\rho}(0) ,
\end{equation}
which we evaluate by exact diagonalization of $-i\hat{\omega}+\hat{R}$,
which is the bottleneck of the calculation since the size of this
matrix grows with $N^4$.

Fortunately the property of the Redfield tensor to couple only
elements of the density matrix $\rho_{mn}$ with
similar energy difference $E_m-E_n \approx M \omega_0$, 
makes the matrix $\hat{R}$ block-diagonal.
The size of each block is only $~4N \times 4N$, which makes the
problem tractable up to hundreds of states in the cavity.

\subsection{Cavity susceptibility}
The cavity susceptibility is defined as
\begin{equation}
\label{CavSusDef}
\chi(\omega)=i \int_0^{\infty} e^{i\omega t} \langle q(t)q(0)-q(0)q(t)\rangle dt,
\end{equation}
where the system is in its steady-state at $t=0$.
In our case the steady state density matrix is diagonal with occupation
numbers, determined by the temperatures of the baths, on the diagonal ($\rho_{steady}$).

Using the quantum fluctuation regression theorem we find
\begin{equation}
\langle q(t)q(0)\rangle = Tr\{q\cdot\mu(t)\},
\end{equation}
where $\mu(t)$ is the solution to the Bloch-Redfield equation
with initial condition $\mu(0)=q \cdot \rho_{steady}$.

Using the solution in Eq.~(\ref{BRsol}) we can perform the
Laplace transform in Eq.~(\ref{CavSusDef}) analytically.
We now note that the main contribution to the susceptibility
around $\omega\approx\omega_0$, which is the main topic of this paper,
comes from the block of $\hat{R}$ connecting elements $\rho_{mn}$ with
energy difference $E_m-E_n \approx \omega_0$.
Thus we only need to diagonalize this single block of size
$~4N \times 4N $ to obtain the susceptibility.

\section{Numerical results}

\begin{figure}
\includegraphics[width=5.5cm]{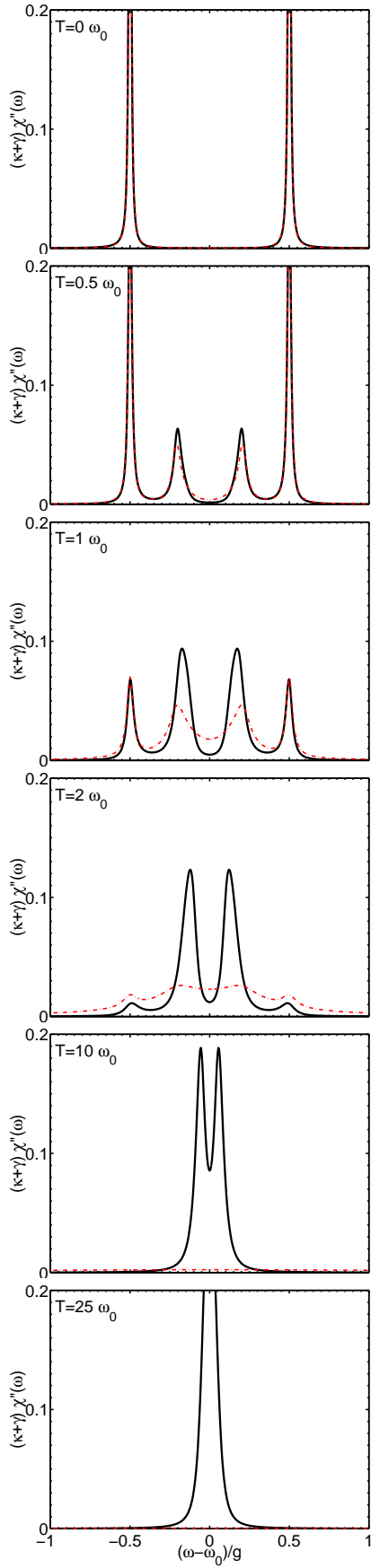}
\caption[]{\label{Fig:NumResAll} Cavity absorption close to the resonance
frequency $\omega_0$, for different temperatures. The paramters are taken
from Ref.~\onlinecite{Blais_Schoelkopf_et_al}, i.e. qubit-cavity coupling 
$g=0.01\omega_0$, cavity quality factor $Q=10^{4}$, and qubit dissipation 
$\gamma=0.08\kappa$. }
\end{figure}

\begin{figure}
\includegraphics[width=5.5cm]{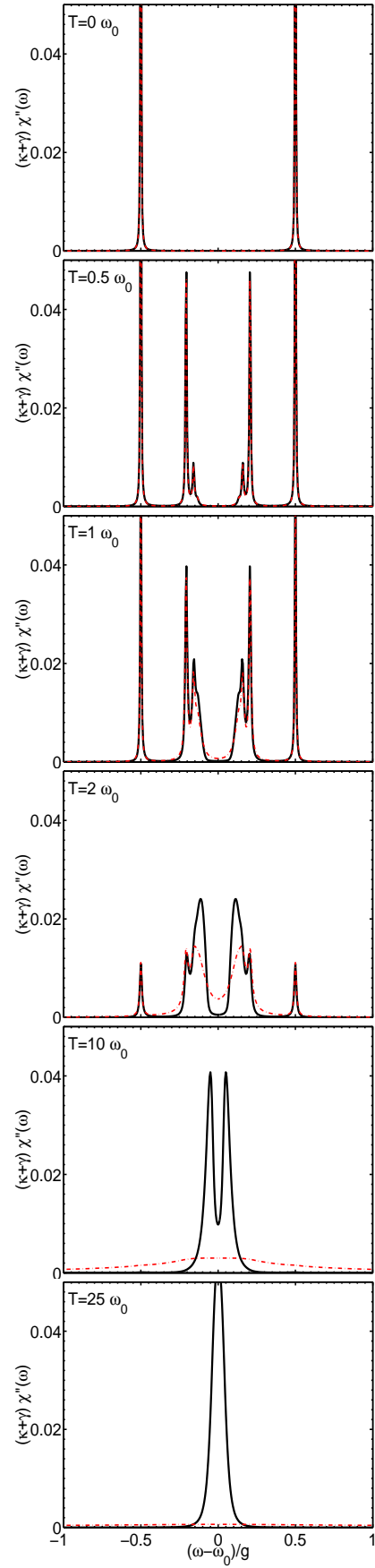}
\caption[]{\label{Fig:Imp_NumResAll} As in Fig.~\ref{Fig:NumResAll}
but with improved quality factor of the cavity $Q=10^{5}$
(implies $\gamma=0.8\kappa$).
}
\end{figure}

The numerical results in Figs.~\ref{Fig:NumResAll}-\ref{Fig:Imp_NumResAll}
show the typical behavior of a Rabi-splitted peak
at zero temperature, the appearance of a broadened
multiple peak structure at intermediate temperatures,
and the merge of the peaks into a single sharp
peak at high temperatures.
The parameters used in Fig.~\ref{Fig:NumResAll} are taken from 
Ref.~\onlinecite{Blais_Schoelkopf_et_al}:
$g=0.01\omega_0$, and $Q=10^{4}$,$\gamma=0.08\kappa$. We assume the
same temperature in both baths.
The dash-dotted (red) line correspond to the secular approximation.

If a higher quality cavity can be achieved, more peaks will be seen before
they merge. This is shown in Fig.~\ref{Fig:Imp_NumResAll},
where $Q=10^5$, and with unchanged qubit dissipation (implies $\gamma=0.8\kappa$).

\section{Conclusions}

In this paper we have studied the regime of cavity QED relevant 
for the Josephson qubits in superconducting cavities. Namely, 
we considered the strong coupling regime when the Rabi splitting 
is much bigger than the inverse life times of the qubit and the cavity,
but the temperature is high. This regime may be realized due to the 
hot photons penetrating the cavity from the manipulating circuits.
We have calculated the susceptibility of the cavity and found that 
it is very sensitive to the temperature. Thus this quantity might 
be used for temperature measurements. On the formal, theoretical 
side we noticed that the secular approximation widely used in the 
applications of the Bloch-Redfield formalism is insufficient when 
Liouvillian degeneracies are present. We compared solutions found
with and without the secular approximation and showed qualitative 
difference between the two at elevated temperatures.

\section{Acknowledgments}

We thank R. Schoelkopf, A. Wallraff, E. Il'ichev, P. Zoller, 
Yu. Makhlin, G. Sch\"on and P. Rabl for fruitful discussions. 
This work is part of CFN (DFG) and was 
supported by the Humboldt Foundation, the BMBF and the ZIP programme 
of the German government, and also by the SQUBIT project of the 
IST-FET programme of the EC, and by Landesstiftung BW. 

\bibliographystyle{apsrev}

\begin{thebibliography}{15}
\expandafter\ifx\csname natexlab\endcsname\relax\def\natexlab#1{#1}\fi
\expandafter\ifx\csname bibnamefont\endcsname\relax
  \def\bibnamefont#1{#1}\fi
\expandafter\ifx\csname bibfnamefont\endcsname\relax
  \def\bibfnamefont#1{#1}\fi
\expandafter\ifx\csname citenamefont\endcsname\relax
  \def\citenamefont#1{#1}\fi
\expandafter\ifx\csname url\endcsname\relax
  \def\url#1{\texttt{#1}}\fi
\expandafter\ifx\csname urlprefix\endcsname\relax\def\urlprefix{URL }\fi
\providecommand{\bibinfo}[2]{#2}
\providecommand{\eprint}[2][]{\url{#2}}

\bibitem[{\citenamefont{Il'ichev et~al.}(2003)\citenamefont{Il'ichev,
  Oukhanski, Izmalkov, Wagner, Grajcar, Meyer, Smirnov, Maassen van den
  Brink, Amin, and Zagoskin}}]{Jena_Rabi}
\bibinfo{author}{\bibfnamefont{E.}~\bibnamefont{Il'ichev}},
  \bibinfo{author}{\bibfnamefont{N.}~\bibnamefont{Oukhanski}},
  \bibinfo{author}{\bibfnamefont{A.}~\bibnamefont{Izmalkov}},
  \bibinfo{author}{\bibfnamefont{T.}~\bibnamefont{Wagner}},
  \bibinfo{author}{\bibfnamefont{M.}~\bibnamefont{Grajcar}},
  \bibinfo{author}{\bibfnamefont{H.-G.} \bibnamefont{Meyer}},
  \bibinfo{author}{\bibfnamefont{A.~Y.} \bibnamefont{Smirnov}},
  \bibinfo{author}{\bibfnamefont{A.} \bibnamefont{Maassen van den Brink}},
  \bibinfo{author}{\bibfnamefont{M.~H.~S.} \bibnamefont{Amin}},
  \bibnamefont{and} \bibinfo{author}{\bibfnamefont{A.~M.}
  \bibnamefont{Zagoskin}}, \bibinfo{journal}{Phys. Rev. Lett.}
  \textbf{\bibinfo{volume}{91}}, \bibinfo{pages}{097906}
  (\bibinfo{year}{2003}).

\bibitem[{\citenamefont{Blais et~al.}(2004)\citenamefont{Blais, Huang,
  Wallraff, Girvin, and Schoelkopf}}]{Blais_Schoelkopf_et_al}
\bibinfo{author}{\bibfnamefont{A.}~\bibnamefont{Blais}},
  \bibinfo{author}{\bibfnamefont{R.~S.} \bibnamefont{Huang}},
  \bibinfo{author}{\bibfnamefont{A.}~\bibnamefont{Wallraff}},
  \bibinfo{author}{\bibfnamefont{S.~M.} \bibnamefont{Girvin}},
  \bibnamefont{and} \bibinfo{author}{\bibfnamefont{R.~J.}
  \bibnamefont{Schoelkopf}}, \bibinfo{journal}{cond-mat/0402216}
  (\bibinfo{year}{2004}).

\bibitem[{\citenamefont{Buisson et~al.}(2003)\citenamefont{Buisson, Balestro,
  Pekola, and Hekking}}]{Buisson_QED}
\bibinfo{author}{\bibfnamefont{O.}~\bibnamefont{Buisson}},
  \bibinfo{author}{\bibfnamefont{F.}~\bibnamefont{Balestro}},
  \bibinfo{author}{\bibfnamefont{J.~P.} \bibnamefont{Pekola}},
  \bibnamefont{and} \bibinfo{author}{\bibfnamefont{F.~W.~J.}
  \bibnamefont{Hekking}}, \bibinfo{journal}{Phys. Rev. Lett.}
  \textbf{\bibinfo{volume}{90}}, \bibinfo{pages}{238304}
  (\bibinfo{year}{2003}).

\bibitem[{\citenamefont{Irish and Schwab}(2003)}]{Irish_Schwab}
\bibinfo{author}{\bibfnamefont{E.~K.} \bibnamefont{Irish}} \bibnamefont{and}
  \bibinfo{author}{\bibfnamefont{K.}~\bibnamefont{Schwab}},
  \bibinfo{journal}{Phys. Rev. B} \textbf{\bibinfo{volume}{68}},
  \bibinfo{pages}{155311} (\bibinfo{year}{2003}).

\bibitem[{\citenamefont{Shore and Knight}(1993)}]{Shore_Knight_Review}
\bibinfo{author}{\bibfnamefont{B.}~\bibnamefont{Shore}} \bibnamefont{and}
  \bibinfo{author}{\bibfnamefont{P.}~\bibnamefont{Knight}},
  \bibinfo{journal}{J. Mod. Opt.} \textbf{\bibinfo{volume}{40}},
  \bibinfo{pages}{1195} (\bibinfo{year}{1993}).

\bibitem[{\citenamefont{Thompson et~al.}(1992)\citenamefont{Thompson, Rempe,
  and Kimble}}]{Kimble_strong_QED}
\bibinfo{author}{\bibfnamefont{R.~J.} \bibnamefont{Thompson}},
  \bibinfo{author}{\bibfnamefont{G.}~\bibnamefont{Rempe}}, \bibnamefont{and}
  \bibinfo{author}{\bibfnamefont{H.~J.} \bibnamefont{Kimble}},
  \bibinfo{journal}{Phys. Rev. Lett.} \textbf{\bibinfo{volume}{68}},
  \bibinfo{pages}{1132} (\bibinfo{year}{1992}).

\bibitem[{\citenamefont{Raimond et~al.}(2001)\citenamefont{Raimond, Brune, and
  Haroche}}]{Raimond_Haroche_RMP}
\bibinfo{author}{\bibfnamefont{J.~M.} \bibnamefont{Raimond}},
  \bibinfo{author}{\bibfnamefont{M.}~\bibnamefont{Brune}}, \bibnamefont{and}
  \bibinfo{author}{\bibfnamefont{S.}~\bibnamefont{Haroche}},
  \bibinfo{journal}{Rev. Mod. Phys.} \textbf{\bibinfo{volume}{73}},
  \bibinfo{pages}{565} (\bibinfo{year}{2001}).

\bibitem[{\citenamefont{Cirac et~al.}(1991)\citenamefont{Cirac, Ritsch, and
  Zoller}}]{Cirac_Dissipative_JCM}
\bibinfo{author}{\bibfnamefont{J.~I.} \bibnamefont{Cirac}},
  \bibinfo{author}{\bibfnamefont{H.}~\bibnamefont{Ritsch}}, \bibnamefont{and}
  \bibinfo{author}{\bibfnamefont{P.}~\bibnamefont{Zoller}},
  \bibinfo{journal}{Phys. Rev. A} \textbf{\bibinfo{volume}{44}},
  \bibinfo{pages}{4541} (\bibinfo{year}{1991}).

\bibitem[{\citenamefont{{Maassen van den Brink} and
  Zagoskin}(2002)}]{Alec_Liouville}
\bibinfo{author}{\bibfnamefont{A.}~\bibnamefont{{Maassen van den Brink}}}
  \bibnamefont{and} \bibinfo{author}{\bibfnamefont{A.~M.}
  \bibnamefont{Zagoskin}}, \bibinfo{journal}{Quant. Inf. Processing}
  \textbf{\bibinfo{volume}{1}}, \bibinfo{pages}{55} (\bibinfo{year}{2002}).

\bibitem[{\citenamefont{Lehmann et~al.}(2003)\citenamefont{Lehmann, Kohler,
  {H\"anggi}, and Nitzan}}]{Haenggi_Rectification_RWA}
\bibinfo{author}{\bibfnamefont{J.}~\bibnamefont{Lehmann}},
  \bibinfo{author}{\bibfnamefont{S.}~\bibnamefont{Kohler}},
  \bibinfo{author}{\bibfnamefont{P.}~\bibnamefont{{H\"anggi}}},
  \bibnamefont{and} \bibinfo{author}{\bibfnamefont{A.}~\bibnamefont{Nitzan}},
  \bibinfo{journal}{J. Chem. Phys.} \textbf{\bibinfo{volume}{118}},
  \bibinfo{pages}{3283} (\bibinfo{year}{2003}).

\bibitem[{\citenamefont{Vion et~al.}(2002)\citenamefont{Vion, Aassime, Cottet,
  Joyez, Pothier, Urbina, Esteve, and Devoret}}]{Saclay_Manipulation_Science}
\bibinfo{author}{\bibfnamefont{D.}~\bibnamefont{Vion}},
  \bibinfo{author}{\bibfnamefont{A.}~\bibnamefont{Aassime}},
  \bibinfo{author}{\bibfnamefont{A.}~\bibnamefont{Cottet}},
  \bibinfo{author}{\bibfnamefont{P.}~\bibnamefont{Joyez}},
  \bibinfo{author}{\bibfnamefont{H.}~\bibnamefont{Pothier}},
  \bibinfo{author}{\bibfnamefont{C.}~\bibnamefont{Urbina}},
  \bibinfo{author}{\bibfnamefont{D.}~\bibnamefont{Esteve}}, \bibnamefont{and}
  \bibinfo{author}{\bibfnamefont{M.~H.} \bibnamefont{Devoret}},
  \bibinfo{journal}{Science} \textbf{\bibinfo{volume}{296}},
  \bibinfo{pages}{886} (\bibinfo{year}{2002}).

\bibitem[{\citenamefont{Feynman and Vernon}(1963)}]{Vernon63}
\bibinfo{author}{\bibfnamefont{R.~P.} \bibnamefont{Feynman}} \bibnamefont{and}
  \bibinfo{author}{\bibfnamefont{F.~L.} \bibnamefont{Vernon}},
  \bibinfo{journal}{Ann. Phys. (NY)} \textbf{\bibinfo{volume}{24}},
  \bibinfo{pages}{118} (\bibinfo{year}{1963}).

\bibitem[{\citenamefont{Caldeira and Leggett}(1981)}]{Caldeira_Leggett_PRL81}
\bibinfo{author}{\bibfnamefont{A.~O.} \bibnamefont{Caldeira}} \bibnamefont{and}
  \bibinfo{author}{\bibfnamefont{A.~J.} \bibnamefont{Leggett}},
  \bibinfo{journal}{Phys. Rev. Lett.} \textbf{\bibinfo{volume}{46}},
  \bibinfo{pages}{211} (\bibinfo{year}{1981}).

\bibitem[{\citenamefont{Bloch}(1957)}]{Bloch_Derivation}
\bibinfo{author}{\bibfnamefont{F.}~\bibnamefont{Bloch}},
  \bibinfo{journal}{Phys. Rev.} \textbf{\bibinfo{volume}{105}},
  \bibinfo{pages}{1206} (\bibinfo{year}{1957}).

\bibitem[{\citenamefont{Redfield}(1957)}]{Redfield_Derivation}
\bibinfo{author}{\bibfnamefont{A.~G.} \bibnamefont{Redfield}},
  \bibinfo{journal}{IBM J. Res. Dev.} \textbf{\bibinfo{volume}{1}},
  \bibinfo{pages}{19} (\bibinfo{year}{1957}).

\end{thebibliography}

\end{document}